\documentclass[%
 reprint,
superscriptaddress,
 amsmath,amssymb,
prl,
]{revtex4-1}
\usepackage[latin1]{inputenc}
\usepackage{graphicx}% Include figure files
\usepackage{dcolumn}% Align table columns on decimal point
\usepackage{bm}% bold math
\usepackage{amssymb}
\usepackage{color}

\begin{document}

%\preprint{APS/123-QED}

\title{Transition from Cassie to Wenzel state in patterned soft elastomer sliding contacts}% Force line breaks with \\

\author{\'Elise Degrandi-Contraires}
\affiliation{Laboratoire de physique des solides, CNRS \& Université Paris-Sud, 91405 Orsay cedex}
\author{Christophe Poulard}
\affiliation{Laboratoire de physique des solides, CNRS \& Université Paris-Sud, 91405 Orsay cedex}
\author{Frédéric Restagno}
\affiliation{Laboratoire de physique des solides, CNRS \& Université Paris-Sud, 91405 Orsay cedex}
\author{Rapha\"el Weil}
\affiliation{Laboratoire de physique des solides, CNRS \& Université Paris-Sud, 91405 Orsay cedex}
\author{Liliane Léger}
\affiliation{Laboratoire de physique des solides, CNRS \& Université Paris-Sud, 91405 Orsay cedex}
\date{\today}% It is always \today, today,

\begin{abstract}

In this paper, we presented an experimental and theoretical analysis of the formation of the contact between a smooth elastomer lens and an elastomer substrate micropatterned with hexagonal arrays of cylindrical pillars. We show using a JKR model coupled with a full description of the deformation of the substrate between the pillars that the transition between the top to the full contact is obtain when the normal load is increased above a well predicted threshold. We have also shown that above the onset of full contact, the evolution of the area of full contact was obeying a simple scaling.

%\begin{description}
%\\pacs  PACS numbers: 81.40.Pq,68.03.Cd
%\end{description}
\end{abstract}

\pacs{Valid PACS appear here}% PACS, the Physics and Astronomy
\maketitle

Roughness is known to deeply influence the contact mechanics of elastic bodies and to drastically affect properties such as adhesion and friction. Due to roughness, only partial contact can usually be established between two solids. The real area of contact is then smaller than the apparent one, and depends on both experimental and material parameters. As a result, molecular forces are usually unable to produce a noticeable adhesion between solids but it has also been shown that dividing a surface in a set of parallel soft asperities can increase adhesion in the case of biomimetical surfaces \cite{Arzt2003}. Concerning friction, roughness is thought to be at the origin of the classical Amonton's law which which predicts that the friction coefficient between two surfaces is independent of the apparent contact area between the surfaces.  In their pionneer work, Fuller and Tabor \cite{Fuller1975} gave the first microscopical understanding of Amonton's law based on the idea that due to plastic deformations of the asperities, their should be a linear relationship between the real contact area between two surfaces and the normal load. More recently, statistical distributions of elastic asperities have been studied by Greenwood \cite{Greenwood67} or Persson \cite{Persson2001a}. Molecular friction is usually studied using SFA ou AFM and is related to monocontact friction but real friction deals with multicontacts. There is however, at present, no real understanding on how one can go from the mono to the multicontact behaviors when changing the roughness or the load, despite the obvious practical importance of being able to adjust and control friction. How does the applied normal load affect the nature of the contact between two solids?  The recent development of microfabrication techniques \cite{McDonald2000} allows a relatively easy preparation of surfaces with well controlled micropatterns having specific geometrical characteristics providing a unique tool to experimentally try to answer to the above questions.
A first exploration of the incidence of micropatterning on sliding friction for elastomeric contacts has been reported recently \cite{Wu2010} and has pointed out the influence of the nature of the contact on the friction between surfaces with controlled asperities made of regular pillars. More precisely, they observed that tall pillars were leading to partial contact, with the contact only established on the top of the pillars, while short pillars were not able to prevent the two surfaces to establish a full contact, due to adhesion forces. The same kind of jump to full contact due to adhesion forces on rough model elastic surfaces has also been studied even more recently on rippled surfaces \cite{Jin2011}.

In the present letter, we present what we think to be the first systematic analysis of the nature of the contact between a smooth elastomeric lens ($R=1.2$~mm and $E_l=2.1$~MPa) and series of elastomeric substrates ($E_s=1.8$~MPa) patterned with regular hexagonal arrays of cylindrical micropillars having a fixed height $h$ and diameters ($h=2.2$~$\mu$m and $d=4$~$\mu$m) and various spacing. This allowed us to put a special emphasis on the incidence of the applied load and of the pattern geometry on the evolution of the nature of the contact. Surfaces were made by moulding and reticulation of PDMS (Sylgard 184) using a know well established technique \cite{Poulard2011}.
%\begin{table}
%  \centering
%  \begin{tabular}{|c|c|c|c|c|c|c|}
%    \hline
%    % after \\: \hline or \cline{col1-col2} \cline{col3-col4} ...
%    $i$ ($\mu$m) & $7$ &$8$ &$9$ &$10$ & $11$ &$12$ \\
%    $\phi$ & $0.296$ &$0.226$ &$0.179$ &$0.145$ &$0.112$ &$0.101$ \\
%    \hline
%  \end{tabular}
%  \caption{Geometrical and mechanical characteristics of the experimental system.}\label{tab_1}
%\end{table}
The homemade JKR apparatus that have been used to investigate the contact have yet been described in \cite{Deruelle1998}. It is devoted to the characterization of the size of the contact versus applied normal load. An essential point is the possibility of continuously monitoring the contact through an optical microscope equipped with a video camera, when progressively pushing to or pulling off the lens from the substrate.
All data reported below correspond to final static state occurring after a micrometric displacement step and a waiting for relaxation.
%All data reported below correspond to final static states when, after a displacement step where the lens is push towards the substrate by a small distance (in the micrometer range), one waits for stationary contact radius and load.

\begin{figure}
  \includegraphics[width=\columnwidth]{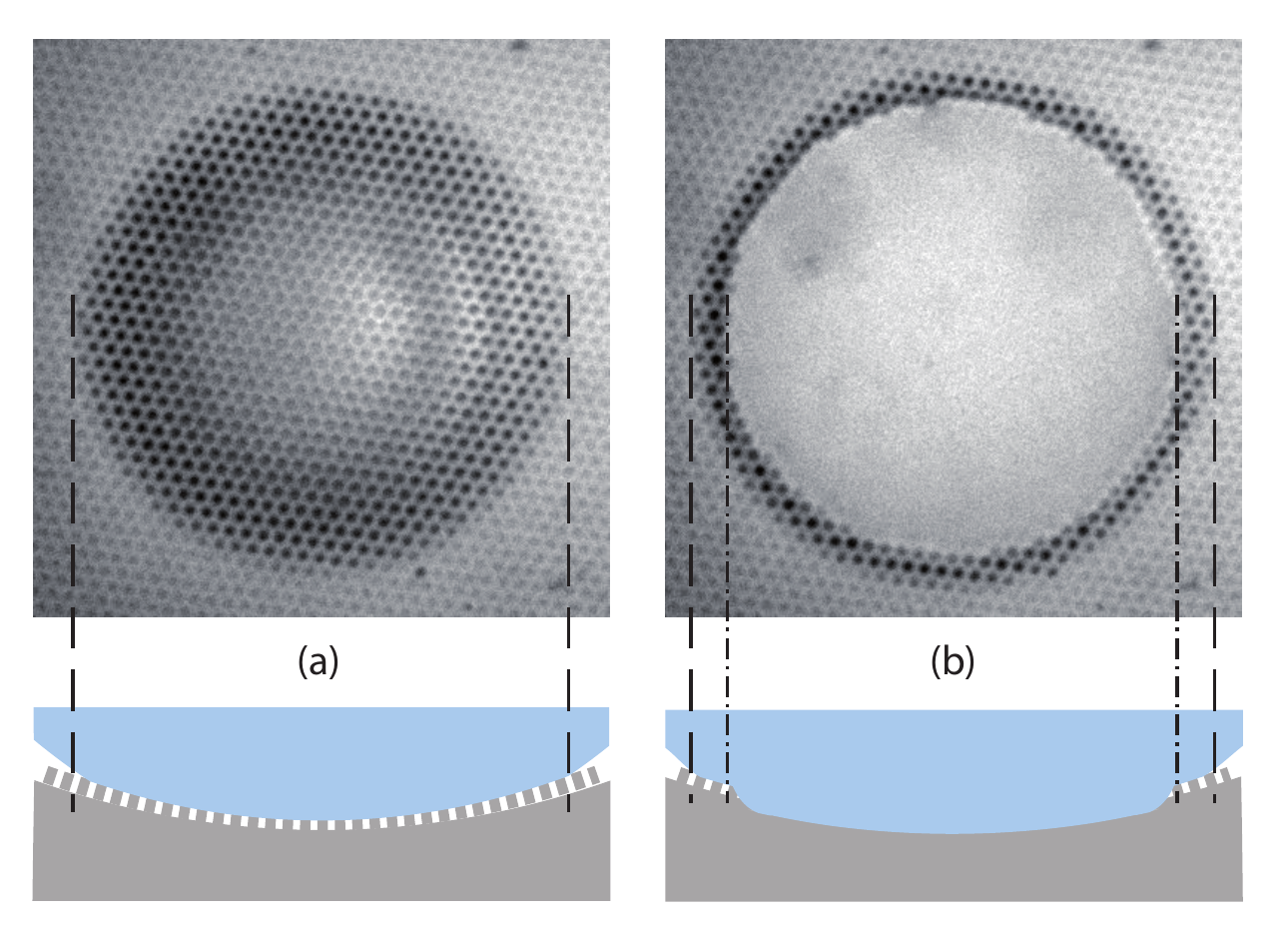}\\
  \caption{Visualization of the contact in top view and schematic representation of the interface for top (a) and mixed (b) contact. }\label{fig1}
\end{figure}

As shown in figure \ref{fig1}, two quite different situations can be identified, depending on the applied load. For low enough loads, the lens only touches the top of the pillars over the whole contact zone (figure \ref{fig1}-a). Air is trapped between the lens and the substrate around the pillars, which then remain well visible due to the index of refraction mismatch between air and elastomer.  For the same pattern geometry, when the normal load is increased, a central zone appears in the contact (figure \ref{fig1}-b), with the lens touching the substrate in between the pillars which then become hardly visible. This zone of full contact remains surrounded by a corona in which the contact is only established on the top of the pillars (contrast similar to that in \ref{fig1}-a). Such a contact will be called a mixed contact.
This transition between top and full contact appears quite reminiscent of the well known Cassie-Wenzel transition observed when a liquid drop is deposited on a rough non wetting substrate \cite{Lafuma2003}.

\begin{figure}
  \includegraphics[width=\columnwidth]{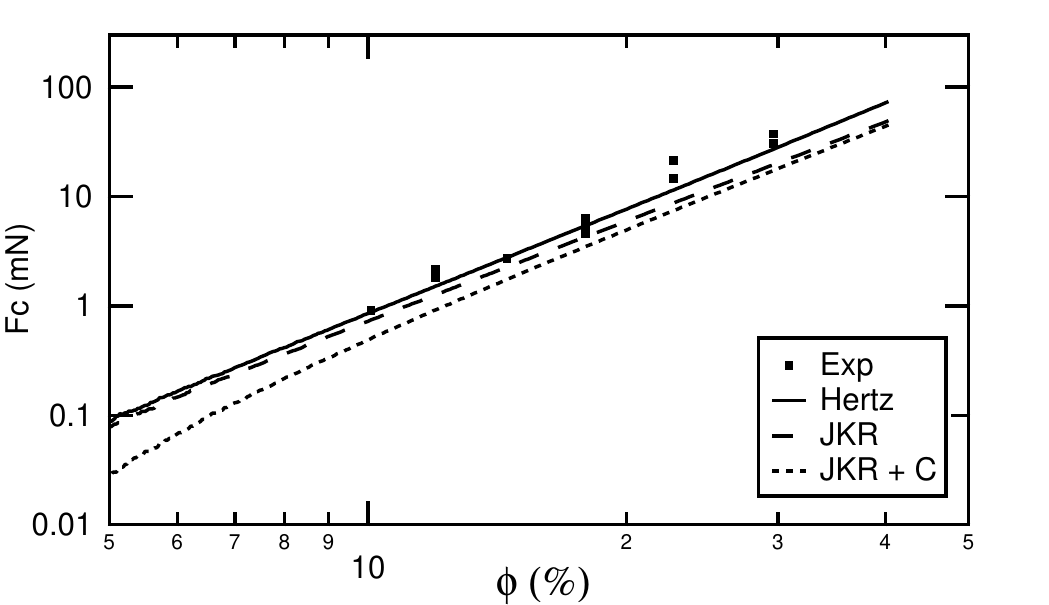}\\
  \caption{log-log representation of the critical normal load versus the fraction of surface occupied by pillars with different spacing.}\label{fig_fig2}
\end{figure}

We have first analyzed in details how the critical load, $F_c$, for the transition between top and mixed contact was affected by the geometrical parameters of the pattern. The results are reported in figure \ref{fig_fig2}, in terms of the evolution of the critical load as a function of the fraction of surface of the substrate occupied by the pillars $\phi$, for pillars all having the same diameter. With the hexagonal array of cylindrical pillars, $\phi=\frac{\pi}{2\sqrt{3}}(d/i)^2$. The range of investigated spacing varying from $5$~$\mu$m to $12$~$\mu$m, $\phi$ was varied from $58$\% to $10$\% (since $d=4$~$\mu$m).
It appears clear in figure \ref{fig_fig2} that the critical force increases with the pillar density $\phi$. Since an increase of $\phi$  means decreasing the pillar spacing $i$, decreasing the distance between pillars makes more and more difficult the formation of a zone of full contact in the center of the contact zone. This is at least qualitatively easy to understand, as, for a given height of the pillars, their relative distance fixes the range over which the lens needs be deformed to accommodate the shape of the patterned substrate.

\begin{figure}
  \includegraphics[width=\columnwidth]{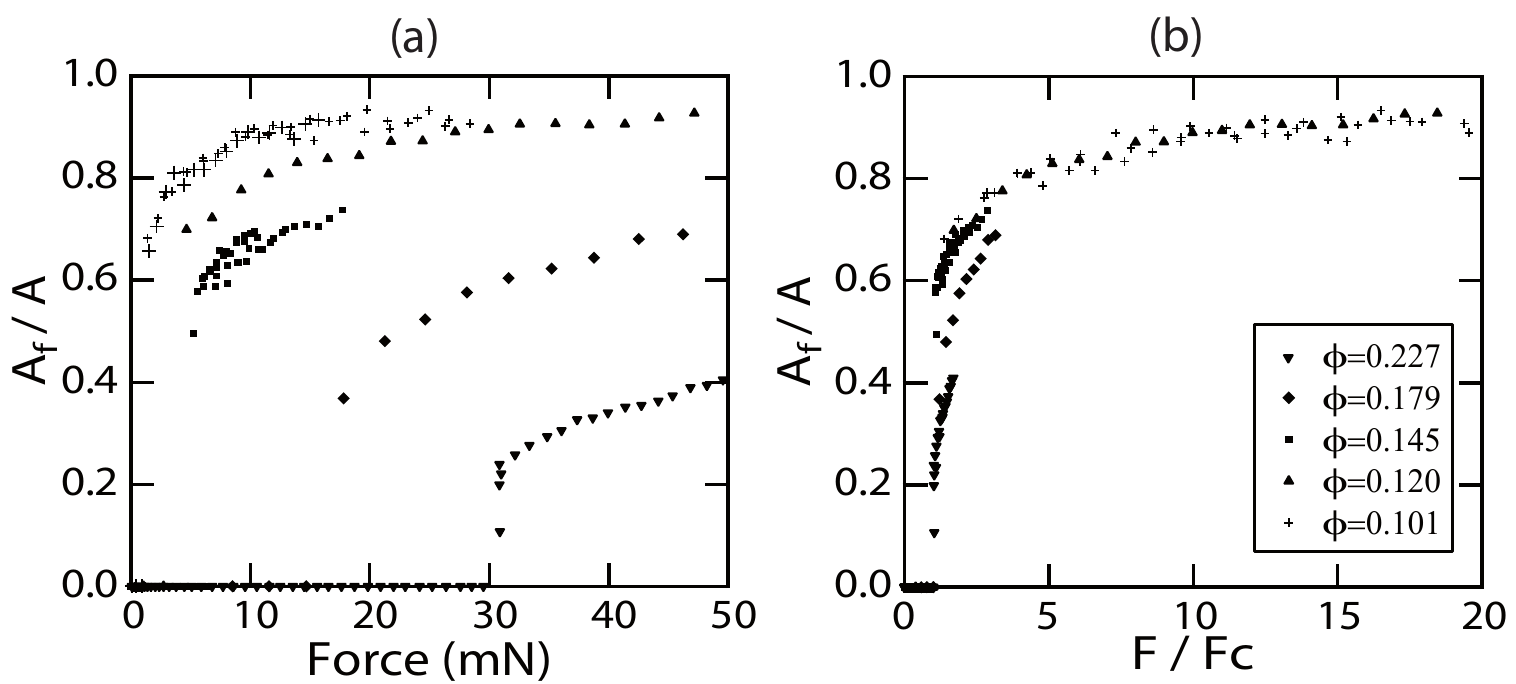}
  \caption{Evolution of the full contact area $A_f$ normalized by the apparent contact area $A$ versus the normal load (left) and the normal load normalized by the critical one (right).
}\label{fig_fig3}
\end{figure}

In figure \ref{fig_fig3}-a, the evolution of the full contact area $A_f$ normalized by the apparent contact area $A$ with the applied normal load is reported for different surface density of pillars. All curves appear rather similar, except for the value of the critical load $F_c$ which depends on $\phi$. It is then tempting to scale all data of the figure \ref{fig_fig3}-a normalizing the applied load by the critical load. Such a scaling is shown in figure \ref{fig_fig3}-b. A single master curve is approximately obtained with an evolution of the scaled area of full contact with the scaled load highly non linear: a rapid increase in the very vicinity of the threshold ($F/F_c\simeq 1$) is followed by a slower long evolution to saturation.
A small deviation is visible to the scaled curve for high surface density of pillars.
The approximate scaling shown in figure \ref{fig_fig3}-b is a remarkable result. We have developed a mechanical description of the contact based on the classical JKR calculation \cite{Johnson1971,Maugis} which allows one to understand the origin of this simple scaling. It point out possible reasons for the departures from scaling well visible in figure \ref{fig_fig3}-b. We present below the main steps of that description.

\begin{figure}
  \includegraphics[width=\columnwidth]{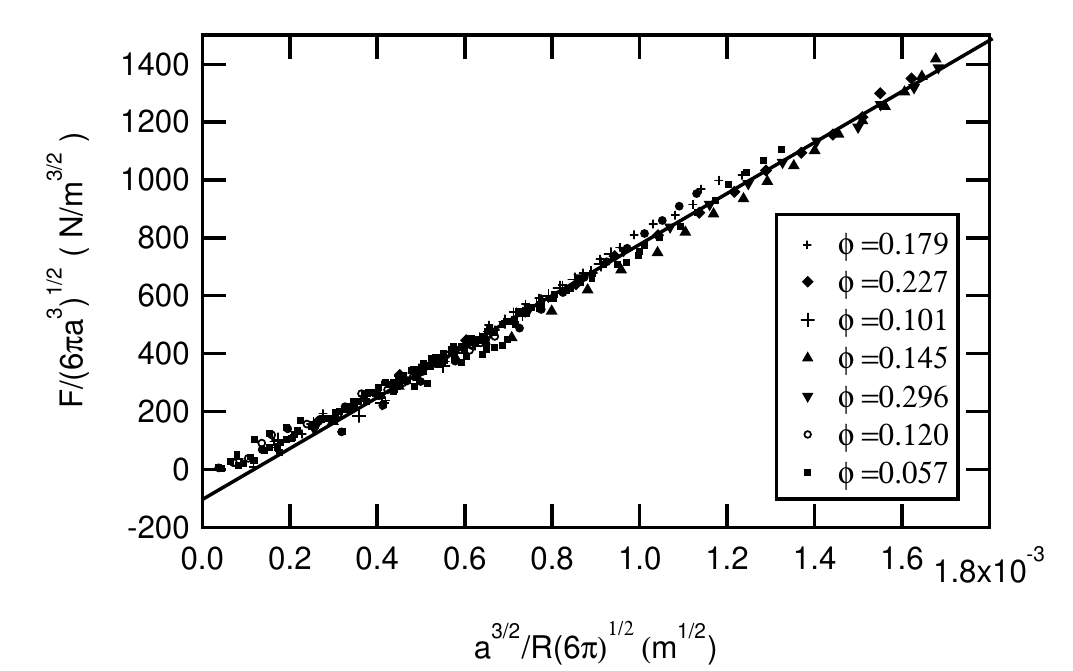}\\
  \caption{
Evolution of the rescaled normal load as a function of the rescaled radius of the contact area for series of patterned surfaces with various patterns characteristics and one elastomer lens with a radius $R=1.2$~mm. }\label{fig_fig4}
\end{figure}

A first important point is to check for the validity of the JKR approach (well known to correctly describe the adhesive smooth contact) for describing either the pure top or the mixed full and top contacts formed in experiments. In figure \ref{fig_fig4}, the relation between the measured normal load (after relaxation at each step) $F$ and the radius of the apparent contact $a$ is reported in the scaled units of the linear form of the JKR equation:
\begin{equation}
\frac{F}{\sqrt{6\pi a^3}}=K\left(\frac{a^{3/2}}{R\sqrt{6\pi}}\right)-\sqrt{W_{\textrm{eff}}K}
\end{equation}
where $R$ is the radius of curvature of the lens and $W_{\textrm{eff}}$ the effective work of adhesion.
All data can clearly be described by a unique linear dependence whatever the nature of the contact (top or mixed) expect for small radius. The slope $K =0.86\pm0.05$~MPa is expected to be the stiffness of the contact. Knowing this value, the intercept at origin allows to define an effective work of adhesion, $W_\text{eff} = 10\pm5$~mJ/m$^2$ having a correct order of magnitude to be identified to  $\phi W$ (with $W$ the work of adhesion on smooth PDMS elastomer, $W = 43$~mJ/m$^2$) for the range of  $\phi$ values span by the experiments ($0.1<\phi <0.3$). One can notice in figure \ref{fig_fig4} a slight tendency to depart from the linear dependence at small effective areas of contact, an effect which can be attributed to finite size effects when, at very small areas of contact, the lens senses essentially the layer of pillars: indeed, in the framework of the overall JKR analysis of the contact, this layer of pillars can be viewed as having an average elastic modulus smaller than a dense smooth elastomer substrate \cite{Barthel2006}.
Two important conclusions can be drawn from the data reported in figure \ref{fig_fig4}: first, the lens - patterned substrate contact globally obeys at the JKR's law, and second, this is true whatever the nature (top or mixed top and full) of the contact.  Globally, the contact can be described through a rigidity modulus, given by those of the PDMS elastomer and lens, and an effective work of adhesion renormalized by the fraction of contact.

As the JKR's law is observed to correctly account for the global mechanics of the contact, the local stress profile inside the contact zone can be given by JKR contact mechanics, i.e. \cite{Maugis}:
\begin{equation}\label{eq_stress_profile}
\sigma=\frac{3f_H}{2\pi a^2}\sqrt{1-\frac{r^2}{a^2}}-\frac{\sqrt{6\pi a^3\phi  W K}}{2\pi a^2}\frac{1}{\sqrt{1-\frac{r^2}{a^2}}}
\end{equation}
where $r$ is the distance from the center of the contact. The maximum local stress is at the center of the contact, which will thus be the location of the maximum deformation of both the lens and the substrate. $F_c$, which is the applied normal force corresponding to the first appearance of a full contact, can thus be obtained by the condition:
\begin{equation}\label{eq_criterion}
\xi_s+\xi_l=h-\delta
\end{equation}
\begin{figure}
  \includegraphics[width=0.8\columnwidth]{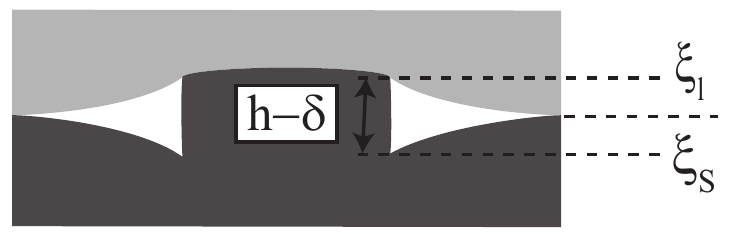}\\
  \caption{Schematic representation of the result of a finite element simulation for the different deformation $\delta$, $\xi_l$ and  $\xi_s$ with periodic boundary condition for $d$=4$~\mu$m and $i$=12$~\mu$m ($\phi=10$\%).}\label{fig_fig5}
\end{figure}
with  $\delta$ the deformation of the pillar located closest to the center of contact,  $\xi_s$  and $\xi_l$  the deformation respectively of the substrate and of the lens, and $h$ the initial height of the pillars. The evaluation of $F_c$ thus amounts to an evaluation of all deformations  $\delta$, $\xi_s$  and $\xi_l$. Considering  $\delta\ll h$, $\xi_l\ll R$ and $\xi_s\ll H$,  the total height of the substrate, all deformations are small and the Hooke's law can be used. One has to notice that, because forces are transmitted through the interfaces, the local stress exerted on one pillar is related to the local stress $\sigma(r)$  given by the JKR stress profile (eq. \ref{eq_stress_profile}) by:
\begin{equation}
\sigma_p=\frac{\sigma(r)}{\phi}
\end{equation}
When the distance between pillars is large enough, both deformations of lens and substrate due to the compression of the more central pillar can be described by the deformation of an incompressible semi-infinite media submitted to a uniform pressure through a flat cylindrical punch, and are given by classical mechanics: $\xi_{s,l}=\frac{2(1-\nu_{s,l}^2)}{\pi}\frac{d\sigma_p}{E_{s,l}}$, with $\nu_{s,l}$  respectively the Poisson ratio and $E_{s,l}$ respectively the elastic modulus for the substrate and the lens. The displacement $\delta$ due to the deformation of a cylindrical pillar under the local stress $\sigma_p$ is: $\delta=\frac{\sigma_p h}{E_s}$.
In fact, as it has been discussed recently, for small $i/d$, the deformation of the substrate (and lens) can no longer relax to the unperturbed position between two close back pillars (coupled behavior of all pillars through the deformation of the underlying substrate \cite{Poulard2011}). The evaluation of the deformations to be plugged in each deformation can then no longer be conducted analytically, but, as shown in \cite{Poulard2011}, both deformation and the stored elastic energy can be estimated numerically. The two deformations of lens and substrate need then to be corrected by a numerically determined correcting function, which is the same on both sides of the interface and only depends on the geometry of the pattern.
In figure \ref{fig_fig2}, the experimental data for the critical force $F_c$ are first estimated using the JKR stress distribution inside the contact, the estimation of the critical force assuming independent pillars (dash line) and coupled pillars (full line) have been plotted. For comparison, the case of independent pillars with a stress distribution given by a Hertz analysis of the contact is also reported as the dotted line. It's important to notice that there are no adjustable parameters in these comparisons. One can see that the Hertz approach is only able to predict qualitatively the observed critical force while, as expected, the JKR mechanics does far better. It is also clear in figure \ref{fig_fig2} that one needs to consider the coupling of the deformations when $\phi$ increases above 0.2 to correctly account for the data.
\begin{figure}
  \includegraphics[width=\columnwidth]{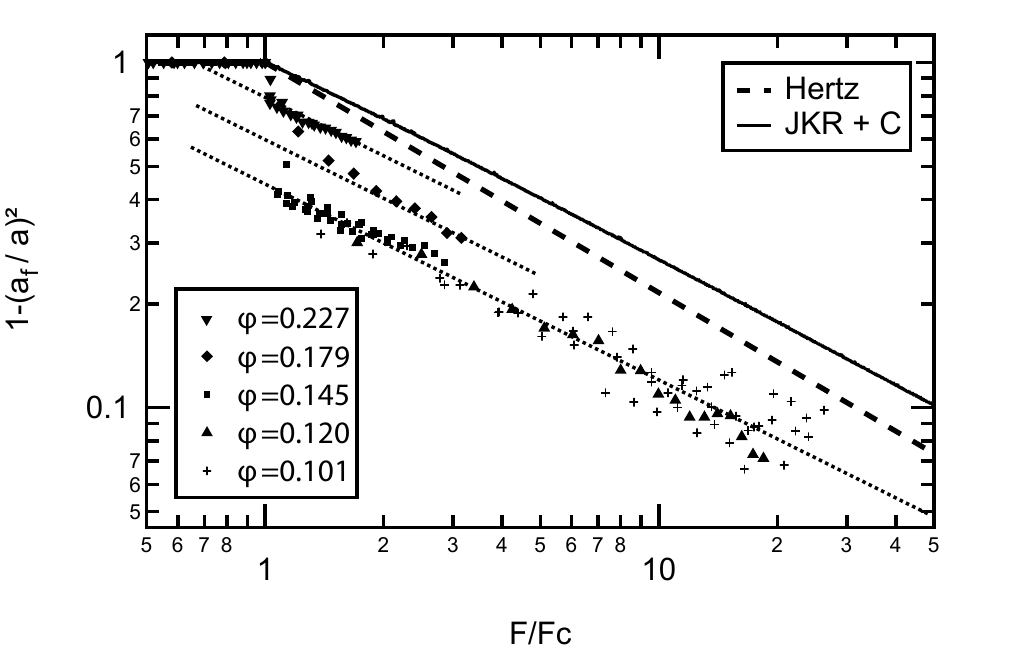}\\
  \caption{Log-log variation for experimental and theoretical values of $(1-(a_f/a)^2)$ versus the normalized force $F/F_c$.}\label{fig_fig6}
\end{figure}

It is now possible to push further the mechanical analysis, in order to try understanding the origin of the approximate scaling shown in figure \ref{fig_fig3}. When increasing the normal force above $F_c$, the radius of the full contact increases and at the border line between top and full contact, the same criterion on deformations as that given in equation  \ref{eq_criterion} holds with now the deformations resulting from the local stress at radius $a_f$, the radius of the full contact. Using in a first approach the Hertz stress distribution $\sigma=\frac{3}{2}\frac{f_H}{(\pi a^2)}\sqrt{1-r^2/a^2}$  and independent pillars lead to a simple scaling defined by:
\begin{equation}
1-\left(\frac{a_f}{a}\right)^2=\left(\frac{F}{F_c}\right)^{-2/3}
\end{equation}
 This is represented as the dashed line in figure \ref{fig_fig6}. The full line represents the result of a similar analysis but using the JKR stress distributions and the numerically estimated coupled pillars deformations (JKR+C). Again, one can clearly see that if the Hertz approach captures the essential features of the problem, it cannot account correctly for the data. In particular, it does not predicts the slope of the linear variation of $\log(1-(a_f/a)^{2})$ versus $\log(F/F_c)$. It also appears clear in figure \ref{fig_fig6} that if the JKR plus coupled pillars approach predicts the exponent of the power law dependence (slope of the dotted lines in the log-log plot of figure \ref{fig_fig6}), it is not able to correctly account for the prefactor. We have right now no real explanation for that fact which may be due to an increase difficulty in increasing the radius of the full contact (a kind of hysteresis of the contact) when the density of pillars is increased.

As a conclusion, we have presented a detailed analysis of the formation of the contact between a smooth elastomer lens and an elastomer substrate micropatterned with hexagonal arrays of cylindrical pillars. We have shown that the transition between top and full contact previously observed in sliding contacts when changing the height of the pillars, could indeed be induced, without sliding motion, when changing the normal load applied in this JKR type contact and studied in a more detail the evolution of the contact area. These findings represent the first step of an extensive investigation of the incidence of patterning in sliding contacts.

\end{document}